\documentclass[11pt,letter]{article}

\usepackage{amsmath}
\usepackage{xspace}
\usepackage{amssymb}
\usepackage{epsfig}
\usepackage{graphicx}
\usepackage{bbold}
\usepackage[usenames,dvipsnames]{color}
\usepackage{cite}

\newtheorem{theorem}{Theorem}
\newtheorem{lemma}[theorem]{Lemma}

\newtheorem{proposition}[theorem]{Proposition}

\newtheorem{observation}{Observation}
\newtheorem{definition}{Definition}

\newtheorem{conjecture}{Conjecture}
\newenvironment{proof}{\noindent {\textbf{Proof }}}{$\Box$ \medskip}

\def\QEDopen{{\setlength{\fboxsep}{0pt}\setlength{\fboxrule}{0.2pt}\fbox{\rule[0pt]{0pt}{1.3ex}\rule[0pt]{1.3ex}{0pt}}}}
\def\QED{\QEDopen}
\def\proof{\noindent{\bf Proof}: }
\def\endproof{\hspace*{\fill}~\QED\par\endtrivlist\unskip}

\newcommand\ket[1]{| #1 \rangle}
\newcommand\bra[1]{\langle #1 |}

\begin{document}

\title{Applications of Monotone Rank to Complexity Theory}

\author{Yang D. Li \footnote{Email: danielliy@gmail.com. University of Illinois at Urbana-Champaign.
}}

\maketitle
\abstract{
Raz's recent result \cite{Raz2010} has rekindled people's interest in the study of \emph{tensor rank}, the generalization of matrix rank to high dimensions, by showing its connections to arithmetic formulas. In this paper, we follow Raz's work and show that \emph{monotone rank}, the monotone variant of tensor rank and matrix rank, has applications in algebraic complexity, quantum computing and communication complexity. A common point of tensor rank and monotone rank is that they are both NP-hard to compute \cite{Has1990, Vav2009}, and are also hard to bound. This paper differs from Raz's paper in that it leverages existing results to show unconditional bounds while Raz's result relies on some assumptions.

More concretely, we show the following things.

\begin{itemize}

\item We show a super-exponential separation between monotone and non-monotone computation in the non-commutative model, and thus provide a strong solution to Nisan's question \cite{Nis1991} in algebraic complexity. More specifically, we exhibit that there exists a homogeneous algebraic function $f$ of degree $d$ ($d$ even) on $n$ variables with the monotone algebraic branching program (ABP) complexity $\Omega(d^2\log n)$ and the non-monotone ABP complexity $O(d^2)$.

\item In Bell's theorem\cite{Bel1964, CHSH1969}, a basic assumption is that players have free will, and under such an assumption, local hidden variable theory still cannot predict the correlations produced by quantum mechanics. Using tools from monotone rank, we show that even if we disallow the players to have free will, local hidden variable theory still cannot predict the correlations produced by quantum mechanics.

\item We generalize the log-rank conjecture \cite{LS1988} in communication complexity to the multiparty case, and prove that for super-polynomial parties, there is a super-polynomial separation between the deterministic communication complexity and the logarithm of the rank of the communication tensor. This means that the log-rank conjecture does not hold in high dimensions.

\end{itemize}
}

\newpage

\section{Introduction}
\emph{Computational complexity} focuses on studying the minimum amount of resources required for carrying out computational tasks. The resources may be time, space, randomness (public or private), communication, quantum entanglement and so on. Readers can refer to the textbook by Arora and Barak \cite{AB2009} for more information on this subject.

\emph{Matrix rank} plays a key role for proving lower bounds, and sometimes upper bounds, in many models of computation with an algebraic or combinatorial flavor, like algebraic branching programs, span programs, and communication complexity. Another important notion is \emph{tensor rank}, which is shown to be crucial in arithmetic formulas \cite{Raz2010}. Tensor rank is the generalization of matrix rank. The monotone variant of tensor rank and matrix rank is called \emph{monotone rank} and is defined as follows, where $[n]$ denotes the set $\{1, 2, \ldots, n\}$, $\otimes$ means tensor product and $\mathbb{R^+}$ represents the set of nonnegative real numbers.

\begin{definition} [monotone rank]
The \emph{monotone rank} of a tensor $M : \Pi_{j=1}^d [n_j] \rightarrow \mathbb{R}^+$ ($d \ge 2$) is the minimum $r$ such that $M = \sum_{i = 1}^r v_{1,i} \otimes v_{2,i} \otimes \cdots \otimes v_{d,i}$, where $v_{j,i} \in (\mathbb{R}^+)^{n_{j}}, j \in [d], i \in [r]$. It is denoted as $mr(M)$.
\end{definition}

When $d \ge 3$, monotone rank is the \emph{monotone tensor rank}, as discussed in \cite{AFT2011}; when $d = 2$, monotone rank specializes to \emph{monotone matrix rank} (or positive rank/nonnegative rank), first mentioned in \cite{Yan1991}. According to a recent report by Lee and Shraibman \cite{LS2009}, there are no lower bounds which actually use monotone rank in practice. The main problem with monotone rank as a complexity measure is that it is extremely difficult to bound for explicit functions. Consequently, to the best of our knowledge, this paper is the first to connect monotone rank to real applications in complexity theory.

A brief preview of our results is that we want to quantify the power of negation. Just like Valiant's famous result \cite{Val1979} on the complexities for computing the permanent and determinant of a matrix,  
the usual and monotone ranks of a nonnegative tensor can also differ dramatically. Morally speaking, the reason for these differences is the obvious possibility of cancelation in the non-monotone case. 
In algebraic complexity, monotone computation does not allow substraction and the coefficients of the monomials are all positive; while the non-monotone model does not have such restrictions. In quantum computing, 
Feynman \cite{Fey1982} points out that the only difference between the probabilistic world and the quantum world is that it happens as if the ``probabilities" would have to go negative in the quantum world. 
Moreover, the separation of communication complexity and the log-rank of the communication tensor is partly due to whether or not we allow negative decomposition.

\subsection{Algebraic Complexity}

Valiant \cite{Val1980} shows an exponential separation between monotone and non-monotone computation in terms of algebraic circuit complexity. Nisan \cite{Nis1991} asks if the same difference between monotone and non-monotone can be achieved in a restricted model called non-commutative model, which prohibits the commutativity of multiplication, for some complexity measure.

We answer this question by showing the following theorem. The gap is more than exponential in terms of algebraic branching program (ABP) complexity, where ABP can be regarded as the analog of branching program in algebraic computation.

\begin{theorem} \label{MT1}
There exists an homogeneous algebraic function $f$ of degree $d$ (d even) on $n$ variables with the monotone ABP complexity $\Omega(d^2\log n)$ and the non-monotone ABP complexity $O(d^2)$.
\end{theorem}

\subsection{Quantum Computing}

Bell's theorem \cite{Bel1964, CHSH1969} basically states that local hidden variable theory cannot predict the correlations produced by quantum mechanics. In Bell's theorem, the measurements are versatile and can be chosen be with respect to various bases; namely the players have the free will to have their own choices and make their own decisions. In a game-theoretic term, they are selfish players. This paper departs from Bell's paradigm by assuming that there are no choices for Alice and Bob and that the measurements Alice and Bob will make are fixed from the start. The model is roughly illustrated by Figure $1$.

\begin{figure*}[t]
\centering
\includegraphics[angle=0, width=0.7\textwidth]{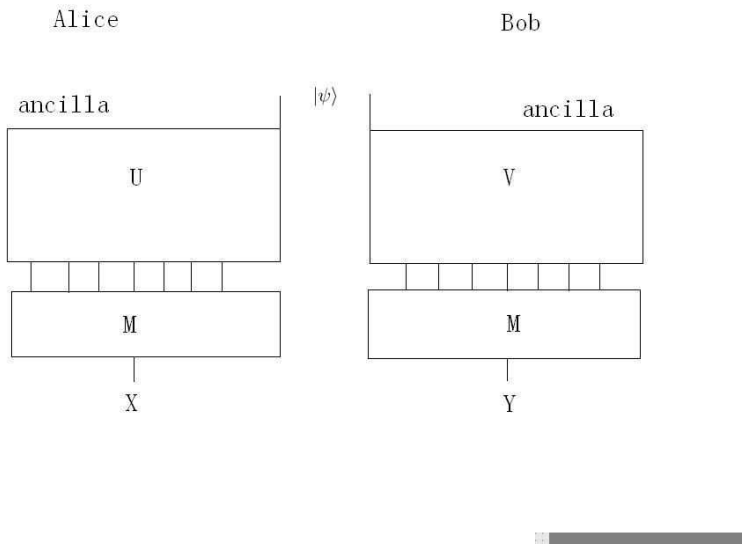}
\vspace{-5pt}
\caption{\label{Figure 1}Bell's Theorem with Fixed Measurement}
\vspace{-5pt}
\label{fig:exp:bra}
\end{figure*}

In Figure $1$, $\ket{\psi}$ represents an entangled state of size $2Q$ ($Q \le n$) qubits, where the first $Q$ qubits are owned by Alice and the second $Q$ qubits belong to Bob. We may also add some ancilla qubits to make sure that both Alice and Bob have $n$ qubits. The state in the beginning is $\phi_0$. Alice applies unitary operation $U$ and Bob applies unitary operation $V$. The state becomes $\phi_1 = (U \otimes V)\phi_0$. $U$ and $V$ are fixed, so that Alice and Bob have no freedom at all. The role of $U$ and $V$ here is to amplify the hardness of simulating quantum correlations.

Then Alice and Bob both apply the measurement $M$, which is fixed to be with respect to the standard basis. $M$ is assumed to be fixed from the start. In the end, they output a correlation $(X,Y)$ according to results of the measurement, where $X$ and $Y$ are random variables taking values in $\{0, 1\}^n$. $X$ and $Y$ are correlated in the sense that their distributions are not independent. Suppose the distribution of $(X, Y)$ is $P_r$, and we want to reproduce $P_r$ using local hidden variables. Note that $P_r$ can be treated as a matrix, namely
\begin{equation}
(P_r)_{xy} = Prob\{X = x, Y = y\},
\end{equation}
for all $x, y \in \{0, 1\}^n$.

For the classical simulation, Alice and Bob initially share some random bits (shared randomness, public coins, or local hidden variables). We denote the shared random variable as $Z$. Alice and Bob also have some private random bits, denoted as $r_A$ and $r_B$ respectively. They use $Z$, $r_A$ and $r_B$ to generate a correlation $(X', Y')$, such that $X' = f_A(Z, r_A)$ and $Y' = f_B(Z, r_B)$. $X'$ and $Y'$ are random variables taking values in $\{0, 1\}^n$. Suppose that the probability distribution of $(X', Y')$ is $P_c$, we want to make sure that $P_c$ is exactly the same as $P_r$. If $P_c$ and $P_r$ are the same, then we succeed in simulating quantum correlations using local hidden variables.

Based on the model above, we have the following result.

\begin{theorem}\label{MT2}
There exists a $2$-qubit ($Q = 1$) quantum state $\ket{\psi}$, some proper $U$ and $V$, such that at least $\log n$ shared random bits are needed to simulate the quantum correlations.
\end{theorem}

Theorem \ref{MT2} is very interesting and subtle: local hidden variables cannot account for quantum correlations even when the measurements Alice and Bob will make are prescribed from the start. That is, Alice and Bob share $Q = 1$ pair and produce $2n$ classical correlated shared bits $(X, Y)$. Local hidden variables fail, not at any particular value of $n$, but in the limit as $n$ tends to infinity, because the number of local hidden variables needed to account for the results grows in a super-exponential speed ($\Omega(\log n)$ vs. $O(1)$) to infinity. Maybe we can say that the problem is that there is no thermodynamic limit. That is to say, the number of hidden variables per pair shared by Alice and Bob is not an intensive quantity. Our argument is not as convincing as Bell's, but it goes beyond Bell's theorem in the sense that it shows that if hidden variables explain quantum correlations, then there is some pathology in the explanation.

\subsection{Communication Complexity}

Arguably the most well-known open problem in communication complexity is the log-rank conjecture \cite{LS1988}, which is stated as follows:

\begin{conjecture}
There exists a constant $c$, such that for every function $f: \{0, 1\}^n \times \{0, 1\}^n \rightarrow \{0, 1\}$,
$$
D(f) = O(\log^c rk(M(f))),
$$
where $D(f)$ is the deterministic communication complexity of $f$ and $M(f)$ is the communication matrix of $f$.
\end{conjecture}

Although a number of people aspire to resolve this conjecture, very little progress has been made in the last two decades \cite{NW1995, RS1995}. In fact, the conjecture can be easily generalized to the number-in-hand multiparty communication complexity model. Suppose there are $d$ parties in the communication.

\begin{conjecture}
There exists a constant $c$, such that for every function $f: (\{0, 1\}^n)^d  \rightarrow \{0, 1\}$,
$$
D(f) = O(\log^c rk(M(f))),
$$
where $M(f)$ is the communication tensor of $f$.
\end{conjecture}

We have the following theorem for the generalized log-rank conjecture.

\begin{theorem} \label{MT3}
For every $d = \omega (n^{c'}), \forall c'>0$, there exists a function $f: (\{0, 1\}^n)^d  \rightarrow \{0, 1\}$, such that for every constant $c > 0$,
$$
D(f) = \omega (\log^c rk(M(f))).
$$
\end{theorem}

Thus, we provide a super-polynomial separation between the deterministic communication complexity and the logarithm of the rank of the communication tensor when there are super-polynomial parties. This means that the log-rank conjecture does not hold in high dimensions.

\subsection{Related Work}
Independent from our paper, in a completely different model and using another complexity measure, \cite{HY2011} shows a much weaker (super-polynomial) separation between the monotone and non-monotone computation in the non-commutative model. We note that \cite{DKW2009} also involves the generalized log-rank conjecture, but it focuses on the case when the number of players is small. There are lots of follow-ups of Bell's theorem. The idea of a great deal of papers (such as \cite{BCT1999, BCvD2001, BT2003, TB2003, RT2009}) is that local hidden variables augmented by communication could reproduce the results of quantum entanglement. Quantum entanglement has plenty of applications in areas such as quantum teleportation \cite{BBCJPW1993}, superdense coding \cite{BW1992} and quantum cryptography \cite{BB1984}. \cite{Zha2012} and \cite{Win2005} are also related in simulating quantum correlations, but they are in a game-theoretic and/or an information-theoretic setting and the selfish players there have the free will. \cite{Hru2011} builds up a relation between monotone rank and the complexity of boolean formula.

\section{Preliminaries}

\subsection{ABP Complexity}

We briefly recall some necessary definitions, notations, and results from Nisan's paper \cite{Nis1991}. For more details, please refer to \cite{Nis1991}. All the logarithms in this paper have base $2$.

\begin{definition}\cite{Nis1991}
An \emph{algebraic branching program (ABP)} is a directed acyclic graph with one source and one sink. The vertices of the graph are partitioned into levels numbered from $0$ to $d$, where edges may only go from level $i$ to level $i+1$. $d$ is called the degree of the ABP. The source is the only vertex at level $0$ and the sink is the only vertex at level $d$. Each edge is labeled with a homogeneous linear function of $x_1, x_2, \ldots,  x_n$ (namely a function of the form $\sum_i c_ix_i$). The size of an ABP is the number of vertices, which is denoted by $B(f)$.
\end{definition}

The ABP model is non-commutative (see \cite{Nis1991}). An ABP computes a function in the following sense: the sum over all paths from the source to the sink, of the product of the linear functions by which the edges of the path are labeled. An ABP of degree $d$ computes a homogeneous polynomial of degree $d$.

\begin{definition}\cite{Nis1991}
An ABP is called monotone if all constants used as coefficients in the linear forms are positive. The monotone ABP complexity of $f$ are denoted as $B^+(f)$.
\end{definition}

\begin{definition}\cite{Nis1991}
For a function $f$ of degree $d$, and $0 \le k \le d$, the $k$-monotone-ABP complexity of $f$, $B_k^+(f)$ is the minimum, over all monotone ABPs that compute $f$, of the size of the $k$'th level of the ABP.
\end{definition}

We use $rk(M)$ to denote the rank of a matrix $M$. Let $f$ be a homogeneous function of degree $d$ on $n$ variables. For each $0 \le k \le d$, we define a real matrix $M_k(f)$ of dimensions $n^k$ by $n^{d-k}$ as follows: there is a row for each sequence of $k$ variables ($k$-term), and a column for each $d-k$-term. The entry at $(x_{i_1}\cdots x_{i_k},x_{j_1}\cdots x_{j_{d-k}})$ is defined to be the real coefficient of the monomial $x_{i_1}\cdots x_{i_k}x_{j_1}\cdots x_{j_{d-k}}$ in $f$.

\begin{lemma}\cite{Nis1991} \label{Thm: upperbound}
For any homogeneous function $f$ of degree $d$, $B(f) = \sum_{k=0}^d rk (M_k(f))$.
\end{lemma}

\begin{lemma}\cite{Nis1991}\label{Thm: directsum}
For every homogeneous function $f$ of degree $d$ and all $0 \le k \le d$, $B_k^+(f) = mr(M_k(f))$. Also, $B^+(f) \ge \sum_{k=0}^dB_k^+(f)$.
\end{lemma}

\subsection{Monotone Rank}

We review some of the known results on monotone rank.

Given $n$ distinct real numbers $a_1, a_2, \ldots , a_n$, a $n \times n$ matrix $M$ can be defined by $M_{ij} = (a_j - a_i)^2$, $i, j \in [n]$. Such a matrix is called a Euclidean distance matrix and has the following properties.

\begin{lemma}\cite{BL2009}  \label{Lem: rank}
$rk(M) = 3$.
\end{lemma}

\begin{lemma}\cite{BL2009} \label{Lem: mrank}
$mr(M) \ge \log n$.
\end{lemma}

\cite{BL2009} conjectures that all Euclidean distance matrices satisfy $mr(M) = n$ and \cite{LC2010a} claims that they prove the conjecture. But Hrubes \cite{Hru2011} refutes the conjecture in \cite{BL2009} and ``theorem" in \cite{LC2010a} by showing the following counter-example.

\begin{lemma}\cite{Hru2011}
Let $a_i = i$ for all $i \in [n]$. Then, $mr(M) \le 2 \log n + 2$.
\end{lemma}

So we put forward a new conjecture based on the new result of Hrubes.

\begin{conjecture}
For all possible distinct real numbers $a_1, a_2, \ldots , a_n$, $mr(M) = \Theta(\log n)$.
\end{conjecture}

A tensor $M : [n]^d \rightarrow \{0,1\}$, which satisfies $M(i_1, i_2, i_3, \ldots, i_d) = 1$ if and only if $\sum_{j = 1}^d i_j $ is divisible by $n$, has the following properties.

\begin{lemma}\cite{AFT2011} \label{Lem: hrank}
$rk(M) \le dn$.
\end{lemma}

\begin{lemma}\cite{AFT2011} \label{Lem: hmrank}
$mr(M) = n^{d - 1}$.
\end{lemma}

\subsection{Hadamard Product}

\begin{definition}
Let $A$ and $B$ be $m \times n$ matrices with entries in $\mathbb{R}$. The \emph{Hadamard product} of $A$
and $B$ is defined by $[A \circ B]_{ij} = [A]_{ij}[B]_{ij}$, for all $1 \le i \le m, 1 \le j \le n$.
\end{definition}

We need a folklore property of Hadamard product.
\begin{proposition} \label{Hadamard}
Let A and B be $m \times n$ real matrices, then $rk(A\circ B) \le rk(A)rk(B)$.
\end{proposition}
\proof
Just note that $A \circ B$ is a submatrix of $A \otimes B$.
\endproof

\section{Monotone vs. Non-monotone Computation}
\subsection{Remarks}
The proof is under the non-commutative model. The method and result do not hold in a commutative setting. In the commutative model the separation between monotone and non-monotone is exponential as shown by Valiant \cite{Val1980} while in our case the separation is super-exponential.
\subsection{Proof of Theorem \ref{MT1}}
Now we prove Theorem \ref{MT1} by defining a function $f$ with ABP complexity $O(d^2)$ and monotone ABP complexity $\Omega(d^2\log n)$. A homogeneous function $f$ of degree $d$ ($d$ even) on $n$ variables is in the form of
\begin{align*}
f(x_1,x_2, \ldots, x_n) = \sum_{i_1, i_2, \ldots, i_d \in [n]} c_{i_1i_2\cdots i_d}x_{i_1}x_{i_2}\cdots x_{i_d}.
\end{align*}
There are totally $n^d$ monomials and $n^d$ corresponding coefficients. We define a function $g: \{i_1, i_2, \ldots, i_{d/2}: i_1, i_2, \ldots, i_{d/2} \in [n]\} \rightarrow [n^{d/2}]$ as follows.

\begin{align*}
g (i_1, i_2, \ldots, i_{d/2}) = \sum_{k=1}^{d/2} (i_k - 1 )n^{d/2 - k} + 1.
\end{align*}

It is easy to verify that $g$ is a bijective function. For instances, $g(1, 1, \ldots, 1, 1) = 1$, $g(1, 1, \ldots, 1, 2) = 2$, $g(1, 1, \ldots, 1, 3) = 3$, ..., $g(n, n, \ldots, n, n - 1) = n^{d/2} - 1$, $g(n, n, \ldots, n, n) = n^{d/2}$.

Then we define $f$ by specifying all its coefficients.

\begin{align*}
c_{i_1i_2\cdots i_d} = (g(i_1, i_2, \ldots, i_{d/2}) - g(i_{d/2 + 1}, i_{d/2 + 2}, \ldots , i_{d}))^2.
\end{align*}

It is clear that all the coefficients are nonnegative. In a word, $f$ is the following.

\begin{align*}
f(x_1,x_2, \ldots , x_n) = \sum_{i_1, i_2, \ldots, i_d \in [n]} (g(i_1, i_2, \ldots, i_{d/2}) - g(i_{d/2 + 1}, i_{d/2 + 2}, \ldots , i_{d}))^2x_{i_1}x_{i_2}\cdots x_{i_d}.
\end{align*}

We arrange $M_{d/2}(f)$ in a way such that $k$-term is in the ascending order of its corresponding $g(i_1, i_2, \ldots, i_{d/2})$, and $d - k$-term is in the ascending order of its corresponding $g(i_{d/2+1}, i_{d/2 + 2}, \ldots , i_{d})$. Then for $M_{d/2}(f)$, it is not hard to verify that $[M_{d/2}(f)]_{ij} = ( j - i )^2, \forall i, j \in [n^{d/2}]$. More explicitly,

\[
M_{d/2}(f) = \begin{bmatrix}
             0^2    &  1^2   &  2^2   &  \cdots  & (n^{d/2} - 1)^2 \\
             1^2    &  0^2   &  1^2   &  \cdots  & (n^{d/2} - 2)^2 \\
             2^2    &  1^2   &  0^2   &  \cdots  & (n^{d/2} - 3)^2 \\
             \vdots & \vdots & \vdots &  \ddots  & \vdots          \\
             (n^{d/2} - 1)^2 & (n^{d/2} - 2)^2 & (n^{d/2} - 3)^2 & \cdots & 0
             \end{bmatrix}.
\]

If we use $R_i$ to represent the $i$-th row of $M_{d/2}(f)$, $M_{d/2}(f)$ could be written into another way.

\[
M_{d/2}(f) = \begin{bmatrix}
             R_1    \\
             R_2    \\
             R_3    \\
             \vdots \\
             R_{n^{d/2}}
             \end{bmatrix}.
\]

More generally, $\forall k \in [d/2]$,

\[
M_{d/2 - k}(f) = \begin{bmatrix}
                 R_1     & R_2     & R_3     & \cdots & R_{n^k} \\
                 R_{n^k+1}& R_{n^k+2}& R_{n^k+3} & \cdots & R_{2n^k}\\
                 \vdots  & \vdots  & \vdots  & \ddots & \vdots \\
                 R_{n^{d/2}- n^k + 1} &  R_{n^{d/2}- n^k + 2} &  R_{n^{d/2}- n^k + 3} & \cdots & R_{n^{d/2}}& \\
                 \end{bmatrix}.
\]

This can be easily verified by definition. For example, when $k = 1$,

\begin{equation}
[M_{d/2 - 1}(f)]_{(x_{i_1}\cdots x_{i_{d/2 - 1}},x_{i_{d/2}}x_{j_1}\cdots x_{j_{d/2}})} = [M_{d/2}(f)]_{(x_{i_1}\cdots x_{i_{d/2}},x_{j_1}\cdots x_{j_{d/2}})},
\end{equation}

and thus,

\[
M_{d/2 - 1}(f) = \begin{bmatrix}
                 R_1     & R_2     & R_3     & \cdots & R_{n} \\
                 R_{n+1}& R_{n+2}& R_{n+3} & \cdots & R_{2n}\\
                 \vdots  & \vdots  & \vdots  & \ddots & \vdots \\
                 R_{n^{d/2}- n + 1} &  R_{n^{d/2}- n + 2} &  R_{n^{d/2}- n + 3} & \cdots & R_{n^{d/2}}& \\
                 \end{bmatrix}.
\]

Next we will show the following two lemmas, the combination of which will immediately yield the result in Theorem \ref{MT1}.

\begin{lemma}\label{ub}
$B(f) = O(d^2)$.
\end{lemma}
\proof
By Lemma \ref{Lem: rank}, $rk(M_{d/2}(f)) = 3$. We define the following subsidiary matrix S of dimension $n^{d/2 - 1}\times (n - 1)$.
\[
S = \begin{bmatrix}
    1^2 & 2^2 & 3^2 & \cdots & (n - 1)^2 \\
    (1 + n)^2 & (2 + n)^2 & (3 + n)^2 & \cdots & (2n - 1)^2 \\
    (1 + 2n)^2 & (2 + 2n)^2 & (3 + 2n)^2 & \cdots & (3n - 1)^2 \\
    \vdots & \vdots & \vdots & \ddots & \vdots \\
    (n^{d/2} - n + 1)^2 & (n^{d/2} - n + 2)^2 & (n^{d/2} - n + 3)^2 & \cdots & (n^{d/2} - 1 )^2
    \end{bmatrix}
\]

A careful comparison of $M_{d/2}(f)$ and $M_{d/2 - 1}(f)$ would reveal the following observation,

\begin{observation}
$rk(M_{d/2 - 1}(f)) \le rk(M_{d/2}(f)) + rk(S)$.
\end{observation}

The correctness of this observation can be very easily verified. Here is a very simple example.

Let $d = 4$ and $n = 2$, then
\[
M_{d/2}(f) = \begin{bmatrix}
             0^2 & 1^2 & 2^2 & 3^2    \\
             1^2 & 0^2 & 1^2 & 2^2    \\
             2^2 & 1^2 & 0^2 & 1^2    \\
             3^2 & 2^2 & 1^2 & 0^2    \\
             \end{bmatrix},
\]

\[
M_{d/2 - 1}(f) = \begin{bmatrix}
             0^2 & 1^2 & 2^2 & 3^2 & 1^2 & 0^2 & 1^2 & 2^2 \\
             2^2 & 1^2 & 0^2 & 1^2 & 3^2 & 2^2 & 1^2 & 0^2   \\
             \end{bmatrix},
\]

and
\[
S = \begin{bmatrix}
             1^2 \\
             3^2 \\
             \end{bmatrix}.
\]

It is not hard to see that the first four columns of $M_{d/2 - 1}(f)$ is a submatrix of $M_{d/2}(f)$. The last four columns of $M_{d/2 - 1}(f)$ differ from the first four columns of $M_{d/2 - 1}(f)$ just by one column, which is $S$. Thus, we know that $rk(M_{d/2 - 1}(f)) \le rk(M_{d/2}(f)) + rk(S)$ for this example.

We define another auxiliary matrix $S_1$.

\[
S_1 = \begin{bmatrix}
    1 & 2 & 3 & \cdots & (n - 1) \\
    (1 + n) & (2 + n) & (3 + n) & \cdots & (2n - 1) \\
    (1 + 2n) & (2 + 2n) & (3 + 2n) & \cdots & (3n - 1) \\
    \vdots & \vdots & \vdots & \ddots & \vdots \\
    (n^{d/2} - n + 1) & (n^{d/2} - n + 2) & (n^{d/2} - n + 3) & \cdots & (n^{d/2} - 1 )
    \end{bmatrix}
\]

It is easy to see that $rk(S_1) = 2$ (Gaussian elimination), and that $S = S_1 \circ S_1$. According to Proposition \ref{Hadamard}, $rk (S) \le (rk (S_1))^2 = 4$. Now we have

$$
rk(M_{d/2 - 1})(f) \le rk(M_{d/2}(f)) + 4.
$$

In a similar way and by a straightforward generalization, we know that
$$
\forall k \in [d/2], rk(M_{d/2 - k}(f)) \le rk(M_{d/2 - k + 1}(f)) + 4.
$$

So we know that $\forall k \in [d/2]$,
$$
rk(M_{d/2 - k}(f)) \le 3 + 4k.
$$
Noting that $M_{d/2 +k}(f)$ and $M_{d/2 - k}(f)$ are symmetric,
$$
\forall k \in [d/2], rk(M_{d/2 +k}(f)) = rk(M_{d/2 - k}(f)) \le 3 + 4k.
$$
Therefore, according to Lemma \ref{Thm: upperbound},

\begin{align*}
B(f) = \sum_{k=0}^d rk (M_k(f)) = O(d^2).
\end{align*}
\endproof

\begin{lemma}\label{lb}
$B^+(f) = \Omega(d^2\log n)$.
\end{lemma}
\proof
By Lemma \ref{Lem: mrank}, we know that $mr(M_{d/2}(f)) \ge \frac{d}{2}\log n$. For $M_{d/2 - 1}(f)$, it is not hard to see that after some permutation of the columns, we can obtain a sub-matrix with dimension $n^{d/2 - 1} \times n^{d/2 - 1}$ from $M_{d/2 - 1}(f)$ as follows:
\[
\begin{bmatrix}
0^2    & n^2       & (2n)^2 & \cdots & (n^{d/2} - n)^2 \\
n^2    & 0^2       & n^2  & \cdots & (n^{d/2} - 2n)^2\\
(2n)^2 & n^2       &  0^2   & \cdots & (n^{d/2} - 3n)^2\\
\vdots & \vdots    & \vdots & \ddots & \vdots          \\
(n^{d/2} - n)^2 & (n^{d/2} - 2n)^2 & (n^{d/2} - 3n)^2 & \cdots & 0^2
\end{bmatrix}.
\]

So $mr(M_{d/2 - 1}(f)) \ge (d/2 - 1)(\log n)$ by Lemma \ref{Lem: mrank}. Similarly, we can show that $\forall k \in [d/2]$,
$$
mr(M_{d/2 - k}(f)) \ge (d/2 - k)(\log n).
$$
Noting the fact that $M_{d/2 - k}(f)$ and $M_{d/2 + k}(f)$ are symmetric, we know that $\forall k \in [d/2]$,
$$
mr(M_{d/2 + k}(f)) = mr(M_{d/2 - k}(f)) \ge (d/2 - k)(\log n).
$$ By Lemma \ref{Thm: directsum},

\begin{align*}
B^+(f)   = \Omega(\sum_{k=0}^dB_k^+(f)) =  \Omega(d^2\log n).
\end{align*}
\endproof

\section{Shared Randomness vs. Quantum Entanglement}

\subsection{Proof of Theorem \ref{MT2}}

In this part, we shall prove Theorem \ref{MT2}. Remember that $U$ and $V$ are unitary matrices of dimension $2^n \times 2^n$, and that $M$ is a measurement with respect to the standard basis. We set $\ket{\psi}$ to be $\frac{\ket{00}+\ket{11}}{\sqrt{2}}$.

First we calculate that $\phi_0 = \frac{1}{\sqrt{2}} (\ket{0^{2n}} + \ket{0^{n - 1}}\ket{11}\ket{0^{n - 1}})$ and $\phi_1 = \frac{1}{\sqrt{2}} (u_0 \otimes v_0 + u_1 \otimes v_1)$,
where $u_0$ is the first column of $U$, $v_0$ is the first column of $V$, $u_1$ is the second column of $U$, and $v_1$ is the $(2^{n - 1} + 1)$-th column of $V$. After the measurement $M$,
there are $4^n$ possibilities. And $Prob\{X = x, Y = y\} =  \frac{1}{2}|u_0(x)v_0(y) + u_1(x)v_1(y)|^2$, for all $x, y\in \{0,1\}^n$. Here we use $x$ and $y$ as the index for vector or matrix.
It is clear that the unitary operation $U$ and $V$ can decide the distribution of the measurement outcome. Suppose $N = 2^n$.
We use a nonnegative matrix $P$ of dimension $N \times N$ to demonstrate the distribution of $(X, Y)$, $P = [Prob\{X = x, Y = y\}]_{xy}$.
Suppose we want to use shared randomness to simulate this distribution generated from quantum entanglement.
In the beginning Alice and Bob share a random variable $Z$, whose sample space is $\Omega$. We would show two lemmas.

\begin{lemma}\label{lowerbound}
$|\Omega| \ge mr(P)$.
\end{lemma}

\begin{lemma}\label{existence}
There exists $U$ and $V$ such that $mr(P) \ge \log N$.
\end{lemma}

From these lemmas, it is clear that $\log |\Omega| \ge \log n$, implying that we need at least $\log n$ bits of shared randomness to simulate the correlation.

\endproof

\subsection{Proof of Lemma \ref{lowerbound}}

We observe that conditional on $Z$, $X$ and $Y$ are independent. That is to say,

\begin{align*}
Prob\{X = x, Y =y\} & = \sum_{z \in \Omega} Prob\{Z = z\} \times Prob\{X = x, Y = y | Z = z\}\\
& = \sum_{z \in \Omega} Prob\{Z = z\} \times Prob\{X = x | Z = z\} \times Prob\{ Y = y | Z = z\}.
\end{align*}

For a fixed $z$, let $v_z$ be the vector of size $2^n$, such that $ v_z(x) = Prob\{X = x | Z = z\}$, and $v_z'$ be the vector of size $2^n$, such that $ v_z'(y) = Prob\{Y = y | Z = z\}$. So,

\begin{align*}
P = \sum_{z \in \Omega} Prob\{Z = z\} (v_z)(v_z')^T,
\end{align*}

which means that $P$ can be decomposed into $|\Omega|$ nonnegative rank-$1$ matrices. By the definition of $mr(P)$, $|\Omega | \ge mr(P)$.

\endproof

\subsection{Proof of Lemma \ref{existence}}

The aim of this lemma is to find a hard distance (the distribution of a quantum correlation) $P$, based on some plausible $U$ and $V$, such that the monotone rank of $P$ is high. Next we will show an explicit $P$ and show why its monotone rank is high.

Let $\{c_x : x \in \{0, 1\}^n\}$ be a set of $N = 2^n$ distinct elements of $\mathbb{R}^+$. and define matrix $C$ to be $C_{xy}=c_y - c_x, x, y \in \{0,1\}^n$. Thus the Hadamard product of $C$ and its conjugate matrix is $C \circ \bar{C} = [(c_y-c_x)^2]_{xy}$. Using Guassian elimination, we know that $rk(C) = 2$. Since $C$ is an antisymmetric matrix, the eigenvalues of $C$ are $\lambda$, $-\lambda$ and $N - 2$  $0$'s. The characteristic polynomial of $C$ is

$$
\sum_{k\in \{0,1...,N\}}e_k \lambda^k,
$$

where $e_k$ is the coefficient of $\lambda^k$. It is easy to see that $e_N = 1, e_{N-1} = 0$ and

$$
e_{N-2} = \sum_{1\le x < y \le N}(c_y - c_x)^2.
$$

For $k \le N-3$,

\begin{align*}
e_k = \sum_{I\subseteq \{0,1\}^n: |I|=N-k} |C_I|,
\end{align*}

where $C_I$ is the submatrix obtained by restricting $C$ on those rows and columns in $I$, and $|C_I|$ stands for the determinant of $C_I$. Since $rk(C) = 2$, $rk(C_I) \le rk(C) = 2$, so $|C_I| = 0$. Consequently, $e_k = 0, \forall k \le N - 3$. Hence, the characteristic polynomial of $C$ is

$$
\lambda^N+ \sum_{1\le x < y \le N}(c_y - c_x)^2\lambda^{N - 2},
$$

and

$$
\lambda = i\sqrt{\sum_{1\le x < y \le N}(c_y - c_x)^2}.
$$

Since $C$ is antisymmetric, $C$ is normal. Using spectral decomposition, we know that

$$
C = \lambda\ket{u_0}\bra{u_0} - \lambda\ket{u_1}\bra{u_1},
$$

where $\ket{u_0}$ is the eigenvector of $\lambda$ and $\ket{u_1}$ is the eigenvector of $-\lambda$. It is easy to take proper distinct values of $\{c_x : x \in \{0, 1\}^n\}$ to satisfy $\sqrt{\sum_{1\le x < y \le n}(c_y - c_x)^2} = \sqrt{1/2}$, which implies $\lambda = i\sqrt{1/2}$. Let $v_0 = \bar{u}_0$ and $v_1 = - \bar{u}_1$ and we get

\begin{align*}
P_{xy} & = \frac{1}{2}|u_0(x)v_0(y) + u_1(x)v_1(y)|^2 \\
& = \frac{1}{2}|u_0(x)\bar{u}_0(y) - u_1(x)\bar{u}_1(y)|^2.
\end{align*}

Also, we have

\begin{align*}
(C \circ \bar{C})_{xy} & = (\lambda u_0(x)\overline{u_0(y)} - \lambda u_1(x)\overline{u_1(y)})\overline{(\lambda u_0(x)\overline{u_0(y)} - \lambda u_1(x)\overline{u_1(y)})} \\
& = \frac{1}{2}|u_0(x)\bar{u}_0(y) - u_1(x)\bar{u}_1(y)|^2.
\end{align*}

Therefore, $P = C \circ \bar{C}$, which means that we are able to construct $P$ by selecting proper values for entries of $C$, where $C$ also determines some columns of $U$ and $V$. By Lemma \ref{Lem: mrank}, $mr(P) = mr(C \circ \bar{C}) \ge \log_2(N)$.
\endproof

\section{Generalized Log-Rank Conjecture}
\subsection{Discussions}

Our communication complexity separation does not hold for randomized communication.

\subsection{Proof of Theorem \ref{MT3}}

A function $f: \{0, 1\}^n  \rightarrow \{0, 1\}$ could be written into an equivalent form $f: [N]  \rightarrow \{0, 1\}$ with $N = 2^n$. For convenience, we will use the latter representation. We define a function $f: [N]^d  \rightarrow \{0, 1\}$ by requiring $f(i_1, i_2, \ldots, i_d) = 1$ if and only if $\sum_{j = 1}^d i_j$ is divisible by $N$. By Lemma \ref{Lem: hrank}, $rk(M(f)) \le dN$. By Lemma \ref{Lem: hmrank}, $mr(M(f)) = N^{d - 1}$. Therefore, $\log rk(M(f)) \le \log d + n$, and $\log mr(M(f)) = (d - 1)n$. Suppose $d = \omega(n^{c'}), \forall c'>0$. Now it is easy to see that for any constant $c > 0$, $\log mr(M(f)) = \omega(\log^c rk(M(f)))$, and because of an obvious relation $\log(rk(M(f))) \le \log(mr(M(f))) \le D(f)$, we know that for any constant $c > 0$, $D(f) = \omega(\log^c rk(M(f)))$.

\endproof

\section{Acknowledgment}
The author would like to thank Boris Alexeev, Michael Forbes, Pavel Hrubes, Eyal Kushilevitz, and Shengyu Zhang for their detailed and helpful comments on an earlier version \cite{Li2011b} of this paper.

\newpage

\bibliography{Monotone_Rank}
\bibliographystyle{plain}

\end{document}